\documentclass[letterpaper]{article}

\usepackage{amsmath,amssymb}

\usepackage{rotating}

\usepackage[font=small,labelfont=bf,
   justification=justified,
   format=plain]{caption} 

\date{}

\usepackage{natbib}
\renewcommand\cite{\citep}

\newcommand\fig[1]{Fig.~\ref{#1}}
\newcommand\sref[1]{Section~\ref{#1}}
\let\oldmarginpar\marginpar
\renewcommand{\marginpar}[1]{\oldmarginpar{\linespread{1}\scriptsize{#1}}}

\newcommand{\noprint}[1]{}
\newcommand{\manufacturer}[1]{}

\begin{document}

\title{Fast Micron-Scale 3D Printing with a Resonant-Scanning Two-Photon Microscope}

\author{Benjamin W.~Pearre (1),\\
Christos Michas (2),\\
Jean-Marc Tsang (2),\\
Timothy J.~Gardner (1,2),\\
Timothy M.~Otchy (1)
\\
((1) Dept of Biology, Boston University, Boston, MA, USA,\\
(2) Dept of Biomedical Engineering, Boston University, Boston, MA, USA)
}

\maketitle

\begin{abstract}
3D printing allows rapid fabrication of complex objects from digital
designs. One 3D-printing process, direct laser writing, polymerises a
light-sensitive material by steering a focused laser beam through the
shape of the object to be created.  The highest-resolution direct
laser writing systems use a femtosecond laser to effect two-photon
polymerisation. The focal (polymerisation) point is steered over the
shape of the desired object with mechanised stages or
galvanometer-controlled mirrors. Here we report a new high-resolution
direct laser writing system that employs a resonant mirror scanner to
achieve a significant increase in printing speed over galvanometer- or
piezo-based methods while maintaining resolution on the order of a
micron. This printer is based on a software modification to a
commerically available resonant-scanning two-photon microscope.  We
demonstrate the complete process chain from hardware configuration and
control software to the printing of objects of approximately
$400\times 400\times 350\;\mu$m, and validate performance
with objective benchmarks. Released under an open-source license, this
work makes micro-scale 3D printing available the large community of
two-photon microscope users, and paves the way toward widespread
availability of precision-printed devices.
\end{abstract}

\noindent  {\bf Comments:} Corresponding author: BWP (bwpearre@bu.edu). TJG and TMO contributed equally to this work.
\vskip 2 mm
\noindent {\bf Conflict-of-Interest statement:} TJG is an employee of Neuralink Inc.
\vskip 2 mm
\noindent {\bf Keywords:} 3d printing, additive manufacturing, lithography, direct laser writing, DLW, two-photon microscopy, resonant scanning.

\section{Introduction}
Direct laser writing (DLW) lithography \cite{maruo1997two-photon-printing} is a
3D-printing technology that can be used to fabricate
small-scale objects with complex geometries by programmatically
exposing a light-sensitive material to a focused laser beam \cite{Atwater2011,
  Buckmann2012, Cumpston1999, Farsari2009,
  Gissibl2016multilens}. Using femtosecond laser pulses and two-photon
polymerisation processes to write a solid object structure into a
photoresist, DLW enables on-demand fabrication
of complex 3D objects with micron-scale features
\cite{Gissibl2016freeform, Malinauskas2010, Malinauskas2013,
  nanoscribe2016, Farsari2009, Kabouraki2015, Gattass2008micromachining, Sun2004, SkylarScott2017VascularNetworks, gottmann_high_2009}. While DLW achieves diffraction-limited resolution, the printing speed of DLW is slow, practically limiting the size of printed objects to millimetres. This speed limitation is changing rapidly, with a number of advancements reported.
The slowest DLW printers use piezo stages, at speeds ranging from $\sim$~0.1--30 millimetres per second \cite{straub_near-infrared_2002,ovsianikov_laser_2011}. Galvanometer-based printers can bring the laser scan rate up to tens or hundreds of mm/s \cite{thiel_direct_2010,maruo_three-dimensional_2000,farsari_two-photon_2006,obata_high-aspect_2013,gottmann_high_2009}. Recent reports include raster-scanned printing with high-speed galvanometers that achieve up to 400~mm/s \cite{SkylarScott2016MicrofabricatedBioscaffolds} by operating the scan mirror near its resonant frequency. Here we extent this trend by incorporating a resonant mirror operating at 8 kHz, which allows printing at speeds up to $\sim 8000$ mm/s.

In order to increase not only the speed of this technology but also its flexibility and availability, we present a raster-scanning DLW (rDLW) system built on a standard resonant-scanning two-photon microscope and open-source control software that are common equipment in many physical and life science laboratories. Open design and
standard commercial components offer easy modification and adaptation to accomodate new materials, object sizes, and techniques.

We demonstrate the capabilities of our resonant rDLW printer in the fabrication of
micron-scale objects. The instantiation reported here is capable of fabricating objects 
that are up to $\sim 400\times 400\times 350\;\mu\textrm{m}$ with minimum feature sizes of $\sim 4\times 1\times 2\;\mu$m (X, Y, and Z, respectively), and with finer X-axis features available through the use of the microscope's zoom. (We note that much larger objects may be 
constructed by stitching together overlapping pieces of this size, although a full 
discussion of this topic is beyond the scope of this work.)
We show that the use of a resonant scanner allows our system to print an object 
of this size and arbitrary geometric complexity in about 20 seconds---a 
significant increase in printing speed over fast galvanometer-based systems.
We evaluate performance with objective metrics assessed using IP-Dip (Nanoscribe, GmbH), a proprietary 
refraction-index--matched resist developed specifically for rapid, 
high-resolution DLW, although we expect that a wide variety of photoresists may be used.

Finally, we release our application software under 
an open-source license. Taken together, this work provides a new platform for innovation in DLW and makes this technology more easily accessible to the community of two-photon microscope users.

\begin{figure}
  \includegraphics[width=\textwidth]{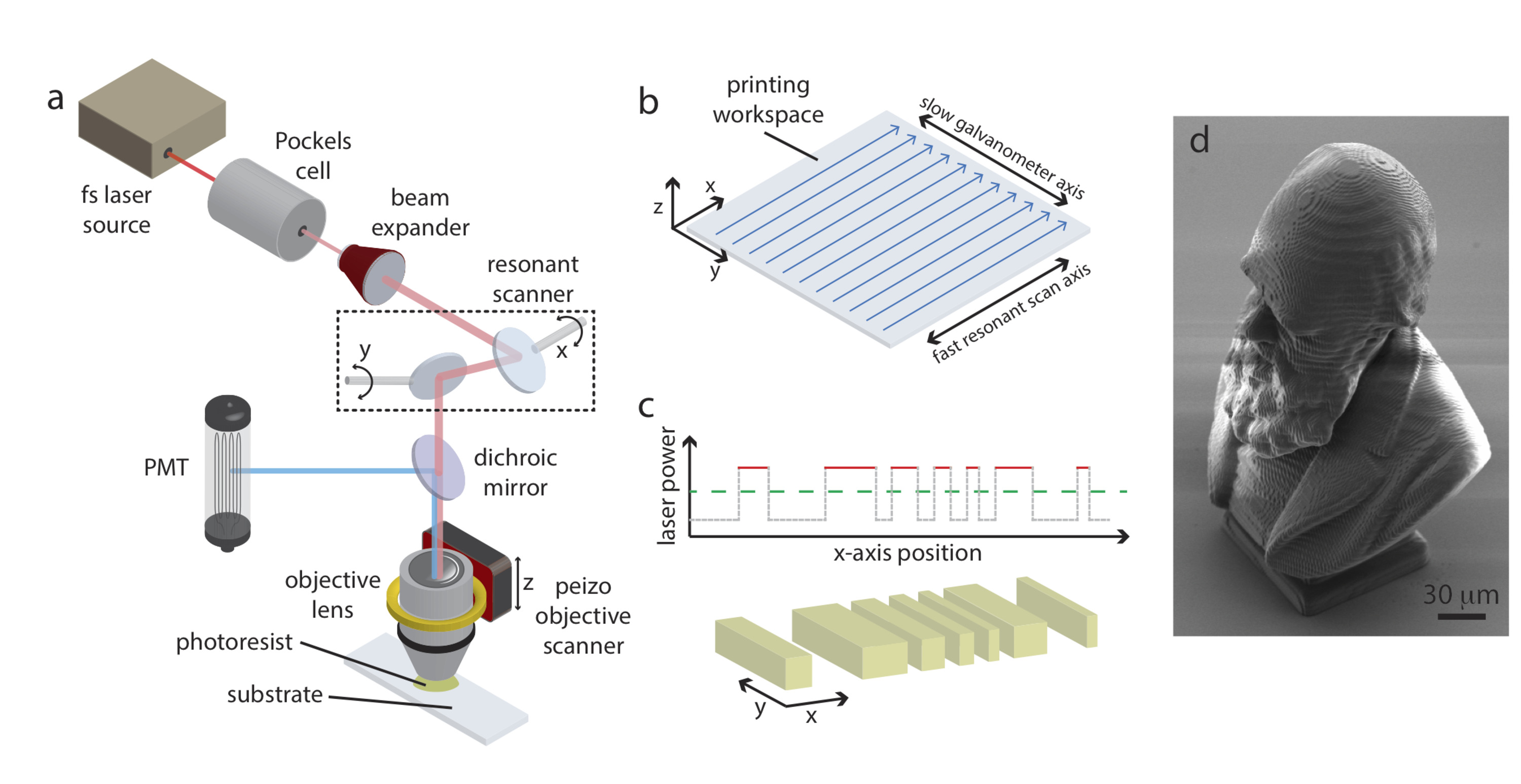}
  \caption{Overview of the resonant rDLW printer. (a) Schematic of the optical
    path from laser source to printed object. (b) The raster scanner
    rapidly sweeps the laser focus across the X axis of the printing
    workspace. (c) Top: laser power is modulated above (red line) and
    below (grey dotted line) the polymerisation threshold (green
    dashed line) throughout the X-axis sweep. Bottom: by applying this
    pattern of laser modulation across the workspace, solid features
    can be built up line by line and layer by layer. (d) SEM
    micrograph of a Charles Darwin statuette printed with our rDLW printer.}
  \label{fig:schematic}
\end{figure}

\section{Results}
The rDLW system we report fabricates objects by
raster-scanning the focal point of a femtosecond laser through a volume of
photoresist, defining the object structure line by line. The device
schematic in \fig{fig:schematic}a depicts the hardware configuration
tested and reported here---essentially, a standard two-photon
microscope with a resonant raster scanner and high-speed/high-extinction-ratio
laser power modulator. This schematic need not be taken as prescriptive, as one of the
benefits to having an open system is the
ability to modify components to meet the requirements of new
applications.

\subsection{A 3D printer built on a two-photon microscope}

Our printer was built around a commercial two-photon microscope platform\manufacturer{Sutter Moveable Objective Microscope}.
A resonant+galvanometer scan module\manufacturer{Sutter MDR-R; 5-mm scan
  mirror} controls the laser's X-Y focal point within the printable
workspace. (Throughout the manuscript, we use Cartesian coordinates to
refer to directions and dimensions in the printing
workspace. Following this nomenclature, X and Y are the perpendicular
axes spanning a single focal plane of the orthogonal Z direction. In
keeping with this, X denotes the direction of the high-speed (7.91 kHz)
raster scanner's sweeps, and Y identifies the slow
galvanometer-controlled row index (\fig{fig:schematic}b)). An immersion objective lens ($25\times$ magnification; numerical aperture (NA) of 0.8) with a refraction compensation
ring\manufacturer{Zeiss LCI Plan-Neofluar; Model
  420852-9972} was used for both printing and imaging. A piezo
scanner\manufacturer{Thorabs PFM450} enabled fast, precise
Z-axis positioning of the objective lens (and hence the focal plane) during
printing. A photomultiplier camera\manufacturer{Hamamatsu
  C8137-02} allowed imaging of the workspace and printed objects.

A tunable Ti-Sapphire laser system\manufacturer{Spectra-Physics
  Millennia Xs pump laser with 3955 Tsunami cavity} ($\sim$120 fs
pulse duration, 80 MHz repetition) provided the light for both
polymerisation and visualisation of the photoresist and printed
objects. We used pump laser powers in the range of 6--10 W, resulting in a $\sim$600--1000-mW mode-locked output beam at the
polymerisation half-wavelength (tunable, but typically 780 nm). Beam intensity was modulated by a Pockels cell (ConOptics 350-80 cell and 302RM voltage amplifier) interfaced with a 3.33-MHz DAC\manufacturer{National Instruments PXIe-6356} (we note that this Pockels cell and driver are not rated for 3-MHz use, but the nonlinearity of the polymerisation reaction allows us to control printing voxelisation at a frequency higher than that for which the Pockels cell is rated. Nonetheless, we recommend that users work with a faster Pockels cell and driver in order to improve small-feature accuracy). Laser intensity was
continuously monitored by sampling the passing
beam\manufacturer{ThorLabs BSF10-B, SM05PD2A, and PDA200C}. To flatten
the profile and improve collimation, the beam was routed through a $2\times$
Galilean beam expander\manufacturer{ThorLabs GBE02-B} before entering
the microscopy optics (\fig{fig:schematic}a).

All components were interfaced with the control computer via a
dedicated data acquisition system\manufacturer{National Instruments
  PXIe-1073 chassis with PXIe-6356 and PXIe-6341 cards}. Vibration due
to floor movements was minimised by building the rDLW system on an
air-shock isolation table\manufacturer{Newport ST Table and I-2000
  Isolators}.

\subsection{PrintImage: a resonant-rDLW control application}

Because the printer is built on a two-photon microscope, we
chose to use a popular open-source microscopy software package,
ScanImage (Vidrio Technologies; Version $\ge$ 5.2.3)
\cite{Pologruto2003ScanImage}, as the basis for system control. To
implement printer functionality, we developed a custom MATLAB
application, PrintImage, that runs alongside ScanImage to control print object
voxelisation, calculate the laser power modulation sequence, and
manage the printing-specific parts of the imaging process.

Print objects may be designed using any computer-aided-design or
engineering (CAD/CAE) software capable of exporting Stereolithography
(STL) files. STL files, which define the unstructured triangulated
surface of the object by the unit normal and vertices of the triangles
using a 3D Cartesian coordinate system, are transformed into a
``watertight'' solid object of specific dimensions that is mapped onto
the predefined set of printer positions via a mesh voxelisation
routine. Voxel Y and Z positions are determined by the number of scan
lines and vertical slices specified by the user; X positions are
computed as described below.

Once the object is voxelised, the series of filled and empty voxels
along the X direction of each Y row (blue arrows,
\fig{fig:schematic}b) is converted into a vector of supra- and
sub-polymerisation-threshold laser powers (\fig{fig:schematic}c) that
defines the geometry (for each Y row) of the printed object. Repeating
this translation for each Y row in every Z plane, the required laser
power at every point within the printer's workspace is computed
before the volume print scan is initiated. Power correction factors (see below) are applied to compensate
for variable beam speed, spherical aberrations in the objective lens,
or other nonuniformities. During printing, ScanImage executes a
volume scan (as is typically performed for volumetric two-photon
calcium imaging) using the laser power sequence precomputed by PrintImage (\fig{fig:schematic}c), thus creating the printed object (\fig{fig:schematic}d).

\subsection{Calibration}
To achieve maximum precision, calibration of imaging and printing
parameters is necessary. We accomplish this by calibrating ScanImage's
optical workspace size parameters, using two methods: (1) producing
fluorescent objects of known dimensions and imaging them with the rDLW
printer, and (2) printing objects with the rDLW printer and measuring
them on a calibrated device.

\begin{figure}
  \includegraphics[width=\textwidth]{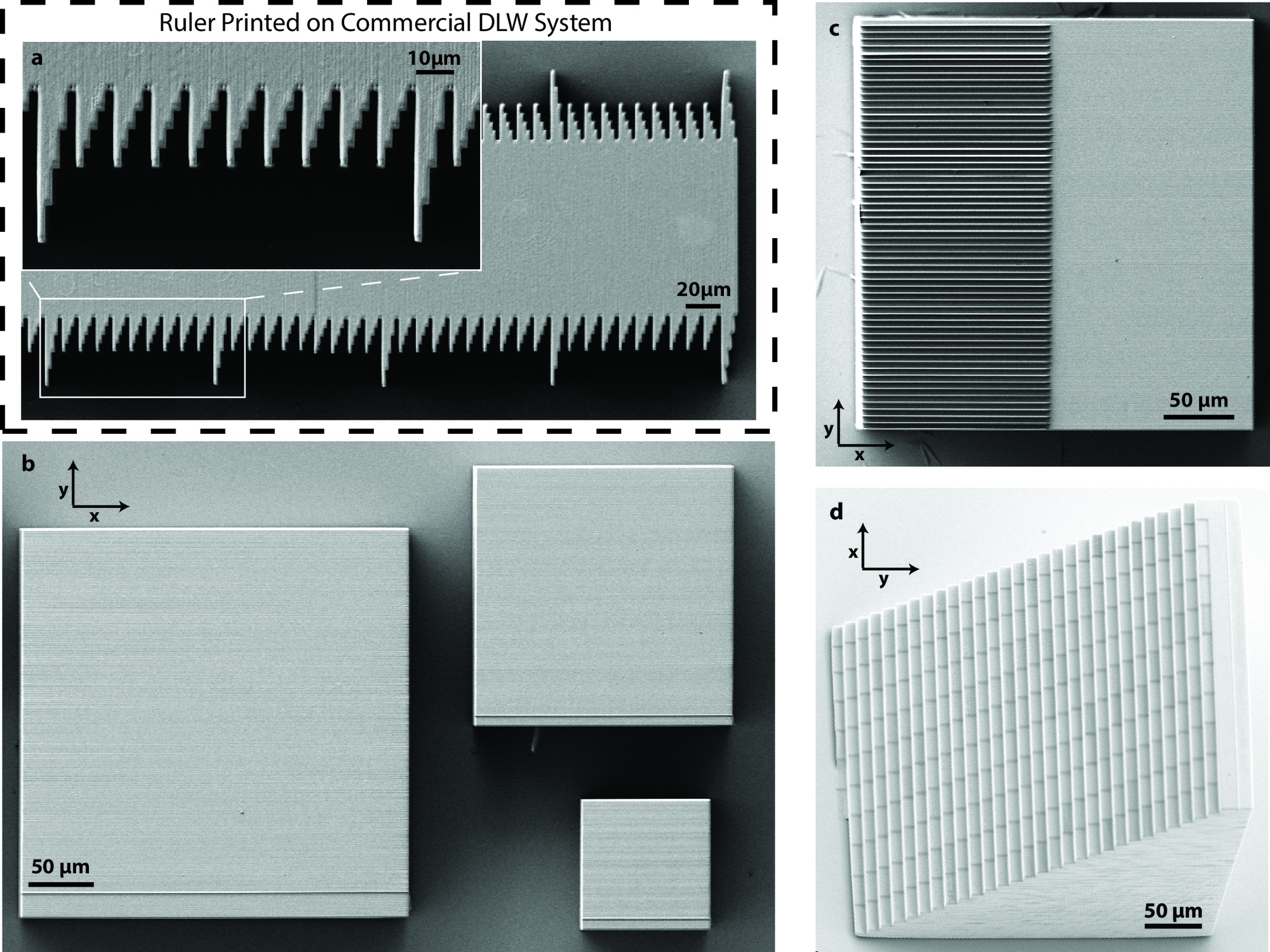}
  \caption{Rulers for calibrating the rDLW system. (a) Ruler
    for measuring X- and Y-workspace dimensions. (b) Cubes used
    to calibrate object size and uniformity of power delivery. The cubes shown have widths 300, 200, and
    100 $\mu$m. The printing parameters were 2.2$\times$ (i.e., 302
    $\times$ 302 $\mu$m FOV), 3.3$\times$ and 6.6$\times$
    magnification (zoom), respectively. Each X-Y plane was built with $152 \times 1024$ voxels,
    and the vertical spacing between the planes was 0.5 $\mu$m for all
    three cubes. (c) Ruler for Y-axis calibration. The printing
    parameters are the same as for the 300-$\mu$m cube in (b). The
    horizontal line spacing on the ruler is 5 $\mu$m.  (d) Vertical
    ruler for Z-axis calibration. Each row along the X axis
    contains 11 steps with 1-$\mu$m height difference. Adjacent steps
    along the Y axis have 10-$\mu$m height difference. The total height
    of the ruler is 300 $\mu$m. The printing parameters were the same
    as for the 300-$\mu$m cube in (b).}
  \label{fig:rulers}
\end{figure}

To create objects of precisely known dimensions, we used a
commercial DLW printer to
print rulers with IP-Dip photoresist (\fig{fig:rulers}a; see
Methods), and confirmed the dimensions of the rulers with SEM
micrographs (the 10-$\mu$m ruler tics measured $9.93\;\mu$m with a SEM two-pixel error $\pm 0.25\;\mu$m). We imaged these rulers with our rDLW printer and adjusted
ScanImage's optical scaling parameters accordingly. We note that 
fluorescent rulers may be created without a calibrated DLW system \cite{Khan2014ruler}.

To calibrate the X-Y plane, we printed calibration cubes
(\fig{fig:rulers}b) and a calibration ruler (\fig{fig:rulers}c) with
the rDLW printer, measured with SEM micrographs the discrepancy between
desired and actual object dimensions, and adjusted ScanImage's workspace size parameters to null the difference. Calibrating the Z
print scale required printing a vertical calibration ruler
(\fig{fig:rulers}d) with regularly spaced Z planes that could be
precisely measured using an optical surface profiler\manufacturer{Zygo NewView
6300}.

\begin{figure}
  \includegraphics[width=\textwidth]{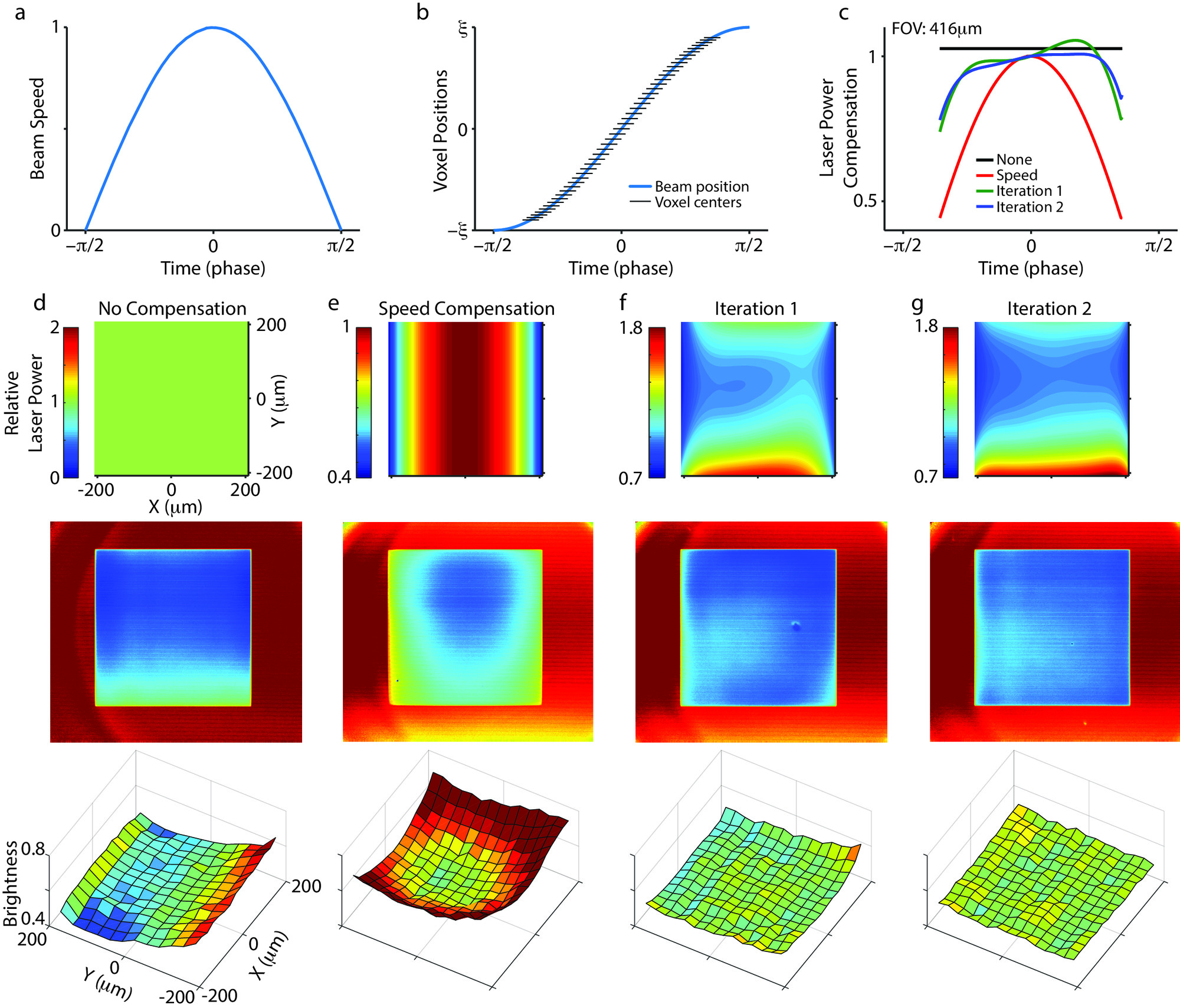}
  \caption{Sinusoidal laser velocity over the X axis results in
    nonuniform voxel size. Both that and optical nonuniformities such as vignetting require corrective
    laser power compensation. (a) Laser focal point velocity as the
    resonant scanner sweeps across the X axis. (b) Laser focus
    position varies sinusoidally with time (blue line). The active
    scanning region is restricted to a portion $D$ of the sweep, with
    X-axis voxel positions shown as black horizontal dashes. For
    clarity, we show where voxels would be defined for an 8-kHz
    resonant scanner with a 1-MHz control system, which yields only 45
    voxels. In order to maintain uniform energy deposition across
    the workspace, laser power is modulated by two factors: it is
    scaled along the X axis by the focal point's speed $\cos (t)$, and
    along the X-Y plane by a learned model of the inverse of optical darkening due to polymerisation. (c) Cross-section of the power compensation along X, in which
    $y=0,\;\;x\in[-208, 208]\;\mu\textrm{m}$ ($1.6\times$ zoom on our rDLW system). (d--g) $400\times 400\times
    100\textrm{-}\mu\textrm{m}$ bricks used to measure and calibrate energy deposition. The upper image shows the print power mask over the
    $208\times 208\textrm{-}\mu$m workspace; the middle image shows an actual printed object (normalised using a baseline fluorescence image); and the bottom image shows brightness data gathered by sweeping the object over the lens so that the same set of pixels in the imaging system may be used for each measurement in order to bypass optical nonuniformities therein. Shown: (d) constant power (note that (1) at this zoom optical vignetting comes close to compensating for X-axis nonuniformity due to varying beam speed; and (2) this image was printed at lower nominal power than the others in order to avoid boiling; for the other images, the speed compensation appropriately reduces power);
    (e) only (X-sinusoidal) speed power compensation; (f--g) two iterations of adaptive power compensation over the visual field (see text). The images and data were obtained with ScanImage on our rDLW system.}
  \label{fig:beamgraph}
\end{figure}

As the resonant scanner sweeps the laser's focal point back and forth across the X axis
of the printer's workspace, the beam moves through the photoresist with
sinusoidally varying velocity (\fig{fig:beamgraph}a). If we
define the centre of the sweep as $t=0$, the oscillation frequency as
$F_r$, and the maximum beam excursion as $\xi$, the focal point's
position $x$ at time $t$ is given by $x = \xi\sin (2 \pi t F_r)$
(\fig{fig:beamgraph}b, blue line). Beam velocity, $\delta x/\delta t$,
rapidly approaches zero at the sweep extremes, so ScanImage restricts the
usable portion of the raster scan to a
central fraction, $D$, of the scan line, resulting in a printing
workspace of width $2\xi D$. From the equation above, one sweep from
$-\xi$ to $\xi$ will take time $t = 1/(2F_r)$, so the beam will traverse the
subsection spanning $D$ in $t = 2\arcsin(D)/(2\pi F_r)$. If laser power (controlled
by the Pockels cell) has a modulation frequency of $F_p$, then power
can be updated every $1/F_p$ seconds; this update rate enables
$r_x=2F_p\arcsin(D)/(2\pi F_r)$ potential changes in laser power level
(i.e., printing voxels) during a single X-axis scan. In our
instantiation, the resonant scanner frequency ($F_r\approx 8$ kHz),
the Pockels cell update rate ($F_p\approx 3.33$ MHz), and ScanImage's
workspace restriction ($D=0.9$), result in a maximum of 152 print
voxels along the X axis.

Resonant-scanner--based control results in higher resolution near the edges of its
sweep than in the centre, but allows higher resolution as workspace size decreases. 
For example, on our rDLW system, printing at 1.3$\times$ zoom
yields a $512\times512\textrm{-}\mu\textrm{m}$ X-Y workspace. On the X axis, voxels are spaced on average every $V/r_x$ for a workspace of span $V$, so at this zoom our rDLW
printer is expected to have a 3.4-$\mu$m mean voxel size along the X axis. At
2.6$\times$ zoom, the mean voxel size along X is expected to be 1.7 $\mu$m over the
$256\times256\textrm{-}\mu\textrm{m}$ workspace.

The use of a resonant scanner leads to significant variation about
this mean, since laser power can be changed only at locations
specified by the position function ($\xi\sin (2 \pi t F_r)$;
\fig{fig:beamgraph}b, black ticks) at a frequency equal to the laser
power modulation rate, $F_p$. Thus, actual voxel size should be 
nonuniform across the X axis, with smaller voxels at the edges of the
workspace than near the center,
proportional to $\cos(\arcsin x)$ for $x\in[-D \xi \ldots D\xi]$
scaled and centred over $V$. As zoom level reduces workspace size $V$, expected voxel sizes over the X axis decrease linearly
until they become limited by optics or photon wavelength (see \sref{sec:resolution}).

\subsection{Varying the laser power to ensure uniform printing}

Given sinusoidally varying scan velocity (\fig{fig:beamgraph}a) and
constant laser power at the focal point, the photoresist will
experience different light exposure conditions as the beam accelerates
from the start of a raster line until it reaches peak velocity at the
centre of the sweep and then deccelerates as it approaches the end of
a line. Under these conditions, the photoresist will not polymerise evenly, and may vaporise or boil in overexposed regions. Thus the
baseline power of the polymerising laser must be corrected by a factor
of $\cos(t)=\cos(\arcsin x)$---proportional to the focal point's
speed---to maintain constant exposure.

Another source of variability in the laser energy available for
polymerisation is attenuation of the beam due to inhomogenieties in
laser intensity over the workspace. This may be due to vignetting,
which attenuates laser power toward the edges of the workspace, or to
other effects such as those resulting from imperfect alignment of
optical components. Falloff due to vignetting is complex, depending on
the angles at which the laser enters and exits each lens in the
system, relative alignments of all optical components, the shape of
the laser beam, partial occlusions throughout the optical path, and
possibly attenuation of the laser beam (although this should be
minimal in its near field). Furthermore, some of these factors may
change frequently in a developing multipurpose tool such as a
two-photon microscope in a research setting.

Due to the difficulty of modeling these factors precisely, we use a
simple adaptive approach to compensate for nonuniform optical
fields. Given a model $M=f:\;x,y\rightarrow$ falloff, power may may be
boosted by $1/M$ to compensate. The liquid photoresist used in these assays (IP-Dip) fluoresces when exposed to
390-nm light (i.e., two near-simultaneous 780-nm photons), and its refractive index and
transparency are functions of the degree of polymerisation. Thus $M$
may be approximated by measuring the reduction in fluorescence of polymerised photoresist
over a uniform printed object (Methods). From these data we fit a
curve such that falloff at any point may be interpolated (in
\fig{fig:beamgraph} we use fourth-order polynomials in X and Y,
although other functions may also be suitable). Due to the nonlinear
relationships between applied laser power and degree of polymerisation
\cite{Mueller2014,Sun2003} and between degree of polymerisation and
reduced fluorescence of the polymerised photoresist, this will not yield a
perfect compensation model in one step, so the process may be iterated
until sufficiently uniform energy deposition is achieved
(\fig{fig:beamgraph}f--g).

We assayed the uniformity of energy deposition across
the workspace by printing $400\times 400\times 100$-$\mu$m solid bricks
and measuring the fluorescence variation across the printed
objects (\fig{fig:beamgraph}d--g) (see Methods). Simple beam speed power
compensation---i.e., reducing power at the extrema of the X axis where the beam moves more slowly---was
effective for producing even power deposition over small objects, but resulted in nonuniformities at
$\lesssim 2.5\times$ zoom: the extreme edges of the X axis fluoresced more brightly than at the centre, indicating a lower degree of
polymerisation.  A compensation function fitted to the measured 
fluorescence variation increased the power at both the X and Y
extrema, compensating for vignetting and other optical irregularities and resulting in nearly uniform polymerisation.

These two forms of power compensation---for focal-point speed and for optical inhomogenieties---are important for uniform
printing, but beyond that they demonstrate the ease with which
polymerisation may be arbitrarily controlled on a per-voxel basis throughout
printed objects, potentially allowing for easy development of
techniques that take advantage of nonuniform polymerisation.

\begin{figure}
  \includegraphics[width=\textwidth]{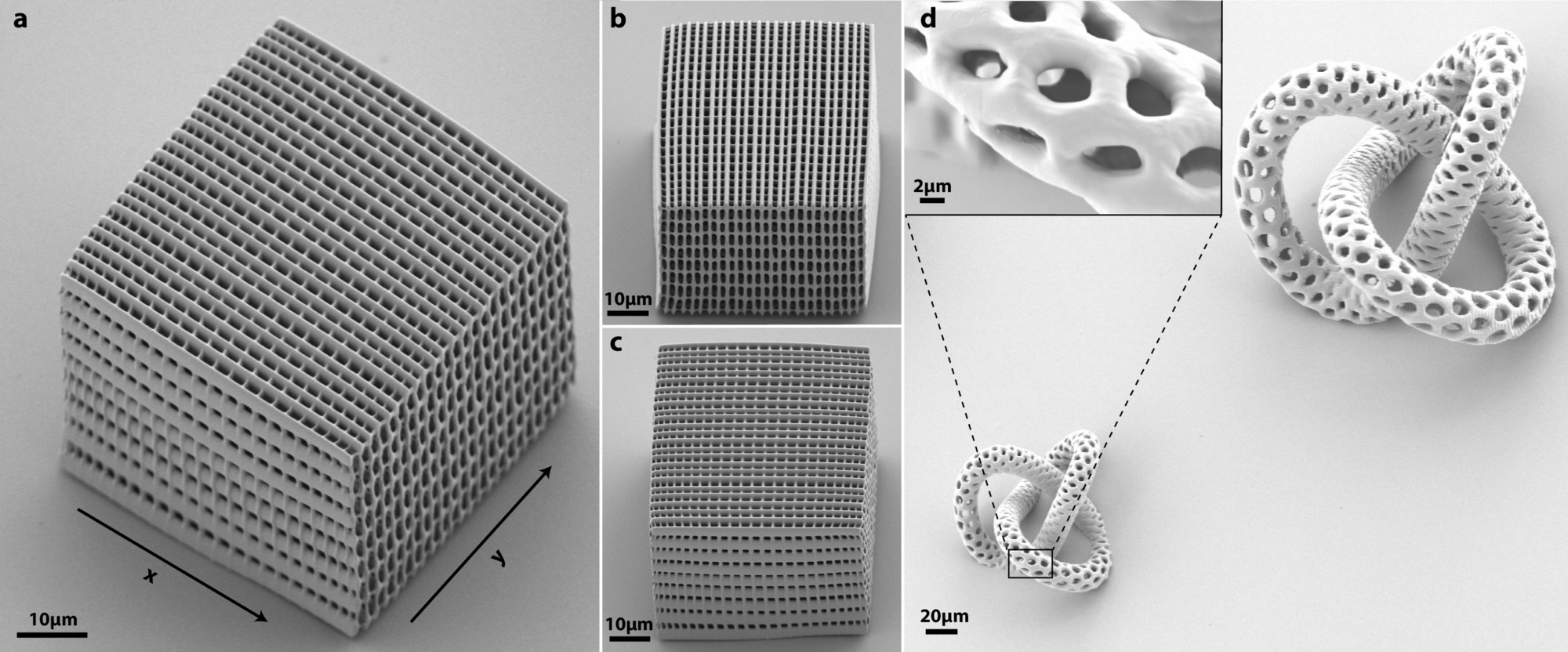}
  \caption{Complex geometric objects printed with our rDLW
    printer. (a-c) Woodpile structure with design dimensions $60\times 60\times 60\;\mu\textrm{m}$. Along
    the X axis, bar thickness was 2 voxels ($0.8\;\mu$m) and bar
    spacing 4 voxels ($1.6\;\mu$m). Bar thickness and spacing on the Y
    axis were 13 and 26 voxels respectively in order to be the same size as the X-axis beams, and
    on the Z axis bars are 1 voxel thick with $6$-$\mu$m spacing. The focal plane
    resolution was $152 \times 1024$ voxels, and the focal plane (Z)
    spacing was $0.2\;\mu$m. (d) A torus knot design
    printed at $100\times 100\times 150\;
    \mu$m (top right) and $50\times 50\times 75\;\mu$m
    (bottom left). The inset shows details within the circumscribed region
    of the bottom left structure. Both knots were printed with focal
    plane resolution $152 \times 512$ voxels and focal plane
    spacing $0.2\; \mu$m.}
  \label{fig:logpile_and_knot}
\end{figure}

\subsection{Accuracy}

We estimated the accuracy of our
rDLW system by printing simple geometric shapes (\fig{fig:rulers}b--d) and comparing
the final object dimensions with those of the original print model (measured with SEM micrographs for X and Y, and the surface profiler for Z). We
found that print errors were not identical across the three dimensions, but instead varied by the print axis. Given
that the laser focal position in 3D space is controlled by three
distinct mechanisms (X axis: resonant scanning mirror; Y axis:
galvanometer mirror; Z axis: piezo objective scanner), and that size is calibrated independently for each dimension, this is
expected. For $300\times 300\textrm{-}\mu\textrm{m}$ cubes printed at $2.2\times$ zoom, we found the errors in the size of the cube to be -$5.6\pm 1.2\;\mu$m ($-1.9\pm 0.4$\%) on the X axis and $6.5\pm 1.0\;\mu$m ($2.2\pm 0.33\%$) on the Y axis ($\pm x$ indicates SEM pixel size $x/2$). Z-axis accuracy was measured using the staircase ruler shown in \fig{fig:rulers}d. Since our printing process leads to small variations (a few microns) in the starting height of the print, we measured Z accuracy at each step of the staircase (we leave more accurate automatic detection of substrate height for future work). The steps had a nominal height of 10 $\mu$m, and an actual mean height of 10.0316 $\mu$m---an error of $\sim 0.32\%$, well within the surface profiler's claimed accuracy of $<0.75\%$.

All measurements were made following immersion of the printed objects in solvent to remove excess/unpolymerised resist (see Methods), and thus our estimates from SEM micrographs include some degree of post-processing--related object shrinkage. Achieving maximum printing accuracy---with this or any other DLW system---requires careful calibration, high-precision components, and control of post-processing deformation. Disentangling the contribution of each to our accuracy estimates is beyond the scope of this work.

\subsection{Resolution}
\label{sec:resolution}
The minimum feature resolution of a two-photon polymerisation process
is a nonlinear function of the precision of laser focal point control,
laser power, and the chemical kinetics of the photoresist
\cite{LaFratta2007}. This complexity makes it challenging to predict
the effective printing resolution of any DLW system, and thus each hardware 
configuration and photoresist combination must be verified experimentally.

One constraint on minimum feature size is the size of the laser's focal point,
as photon density is sufficiently high to initiate the polymerising
reaction throughout this region. The laser focal point radial
(i.e., along the X- and Y-axes) and axial (along the Z-axis)
dimensions are functions of the laser's wavelength, $\lambda$, and the
numerical aperture, NA, of the objective lens. Assuming an ideal
(i.e., Gaussian) beam profile, the full-width half-maximum size of the
point spread function is $\lambda / (2\cdot \textrm{NA})$ (radial) and
$\lambda / (2\cdot \textrm{NA}^2)$ (axial) \cite{Urey2004}. With our
operating wavelength (780) nm and objective NA (0.8), the
theoretical focal point dimensions are 488 and 609 nm,
respectively. Other factors affect the effective size of the focal
point---for example, if the laser beam incompletely fills the back of
the objective, the effective NA will be lower, whereas changing the
laser's power will control the portion of the point spread function
that crosses the polymerisation threshold \cite{Kawata2001submicron}.

\begin{figure}
  \includegraphics[width=\textwidth]{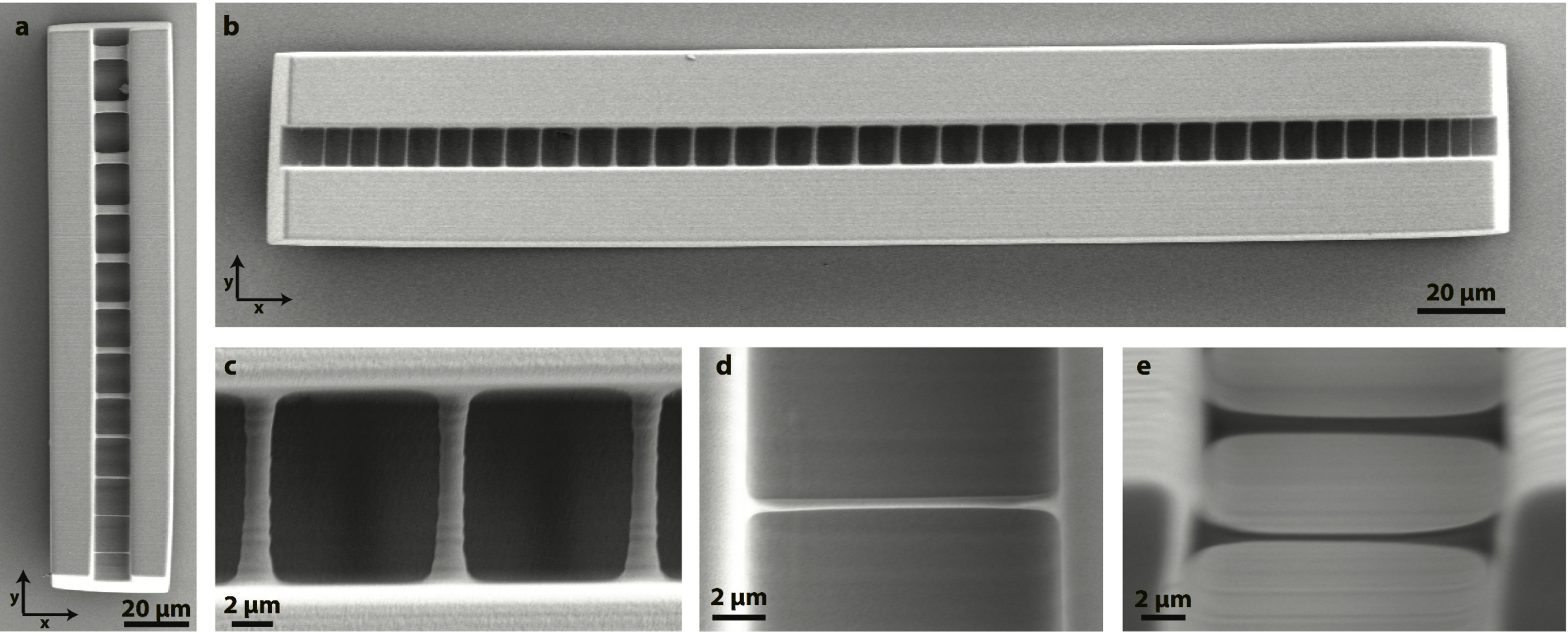}
  \caption{Objects with one- and two-voxel features printed on our rDLW system. (a) Object
    used to estimate minimum voxel size on the Y and Z axes. All
    bridges have single-voxel height (Z), and increasing width on the
    Y axis. The bottom bridge has one-voxel width; thus, it gives an
    idea of the thinnest suspended structure that can achieved with
    the used parameters and photoresist. The object was printed with
    2.2$\times$ magnification for a 302$\times$302 $\mu$m field of
    view. The resolution of each focal plane is $152 \times 1024$
    voxels, and the vertical distance between Z planes is 0.5
    $\mu$m. (b) Object used to estimate the voxel size on the X
    axis. The printing parameters are the same as in (a). The
    bridges were designed to be two voxels wide on the X axis, so
    their size follows a sinusoidal distribution due to the
    cosinusoidal speed profile of the laser beam. (c) Top view of the
    central bridge of (b), which represents the largest value in
    the workspace of double-voxel X resolution at this zoom level.
    (d) Top view of the lowest bridge of (a). (e) View of the lowest bridge of (a)
    at 60$^{\circ}$ from the top view.}
  \label{fig:resolution}
\end{figure}

We treat the location of the centre of the focal point as the voxel
location, since the degree to which it can be controlled defines
another constraint on feature size.
While the Y and Z positioning of the focal point are addressable with
sub-micron accuracy via analogue control of the galvanometer mirror and
the objective-lens scanner, respectively, the continuous sinusoidal
motion of the resonant scanline along the X axis precludes direct
control of position. Instead, X-axis voxel positions and sizes are defined by
the rate at which the laser beam power can be modulated across
the polymerisation threshold. Given the Pockels cell update frequency and the
resonant scanner sweep rate, we estimate the X-axis
resolution to be $\sim$ 2.5--5.6 $\mu$m (at the edge and centre of the resonant sweep, respectively) at 1$\times$ zoom, with minimum feature size decreasing linearly with increasing magnification---for example, the 300-$\mu$m scale used for the resolution tests in \fig{fig:resolution} should have X voxel widths of $\sim$ 1.1--2.5 $\mu$m (edge and centre, respectively). In the following discussion we report only worst-case resolution---that at the centre of the X sweep.

Given these theoretical estimates of printer performance, we
quantified real performance by printing objects with thin single-voxel
features (\fig{fig:resolution}a,b) and measuring feature
dimensions using SEM micrographs. During polymerisation, the forming
polymer tends to shrink due to both the emerging binding forces and
polymerisation quenching from molecules present in the liquid
solution. These effects are reduced when the forming polymer is
attached to a solid object (e.g., a previously polymerised
structure). We report the feature size as it is measured in features
attached to a larger solid structure, but we also note the sizes of
isolated features. We printed structures that consisted of channels
containing single-voxel bridges, at 2.2$\times$ zoom (an FOV of
$302\times 302\;\mu$m). Because single-X-voxel bridges often broke
during postprocessing, we also report 2-X-voxel--wide bridges. We estimated resolution by measuring the size of
the bridges where they
are attached to the supporting walls, and also estimated minimum size of suspended features---giving an idea of possible shrinkage---by measuring the bridges at their centre. Results are shown in Table~\ref{table:resolution}.
\begin{table}[t]
 \begin{center}
   \begin{tabular}{c|ccc}
     & Voxel & Attached & Isolated \\
     & dimension ($\mu$m) & ($\mu$m) & ($\mu$m) \\
     \hline
     X & 2.5/1.1 & 1.09\,$\pm$0.14 & 0.3\,$\pm$0.14 \\
     2X & 5.0/2.2 & 2.73\,$\pm$0.06 & 1.26\,$\pm$0.06 \\
     Y & 0.3 & 1.25\,$\pm$0.03 & 0.35\,$\pm$0.03 \\
     Z & 0.5 & 2.10\,$\pm$0.05 & 0.45\,$\pm$0.05
   \end{tabular}
 \end{center}
 \caption{Printing resolution estimates. Resolution was estimated from
   SEM micrographs of the single- and double-voxel bridges in the
   objects shown in \fig{fig:resolution}.  Measurements of attached features were
   made proximal to the wall of the support structure;
   isolated-feature sizes were measured at the bridge centres.
   ``Voxel dimension'' is defined by the cell sizes used for voxelisation. We list theoretical voxel dimension for the X axis as two numbers: at
   the centre and edges of the resonant scanner's sweep, respectively.
   We report measurements of X feature size at the centre of the resonant sweep---the region of the workspace in which we expect the largest minimum feature sizes. Discrepancies are expected due to the nonzero size and anisotropic shape of 
   the focal point, postprocessing deformation, and a possible difference between rise and fall response times of our Pockels system.}
 \label{table:resolution}
\end{table}

The reported resolutions can be used to build complex
thin-feature structures, such as the 60-$\mu$m woodpile shown in
\fig{fig:logpile_and_knot}a--c and the hollow torus knot structure in
\fig{fig:logpile_and_knot}d.
Though these minimum feature sizes will be sufficient for some
applications, further improvement is possible. As the X-axis voxel
resolution is in part defined by the rate of laser power modulation,
upgrading the Pockels cell and/or control hardware may produce a
substantial reduction in feature size across the whole X axis. Improvement
in all axes could be achieved by reducing the focal point
size (e.g., by increasing the objective lens aperture, flattening
the beam profile, or reducing power \cite{Kawata2001submicron}), or using photoresists
with higher polymerisation thresholds or reduced spatial expansion
factors.

\subsection{Speed}
A key aim for our rDLW design was to increase
fabrication speed through the introduction of a resonant scanner.
As a first approximation, fabrication time is governed by two
parameters: the speed with which the beam moves through the resist and the linear
distance that the beam must traverse \cite{Malinauskas2013}. DLW systems typically use 
some combination of stage-based (i.e., using motorised stages to move the printed object 
relative to a stationary laser focus) and mirror-based (i.e., using mirrors to move the laser 
focus relative to a objective's stationary object) methods for polymerising the desired location. Each has its advantages, making direct comparisons challenging, but 
mirror-based scanning is capable of realising significantly higher scanning speeds while
maintaining micro- and nanoscale feature sizes (Table~\ref{table:speed}). 

Many DLW systems 
realise significant time savings by optimising the laser
path such that travel distance is minimised. For printed objects
with small fill ratios, this strategy can produce substantial
improvements in fabrication speed. Other strategies, such as the core-and-shell 
printing process \cite{nanoscribe2016shell}, can reduce 
fabrication time for objects with low surface-area/volume ratios. Our approach achieves uniform 
fabrication times across fill ratios by using a resonant scanner to sweep the 
beam over every point in the printing workspace (\fig{fig:schematic}b), maximising travel
distance but at a higher mean speed than in many previously described systems 
(Table~\ref{table:speed}).

\begin{table}[t]
	\begin{center}
		\begin{tabular}{c|ccc}
			Positioning & Scanning & Nominal Feature & \\
			Mechanism & Speed (mm/s) & Size ($\mu$m) & Reference \\
			\hline
			Stepper motor & 10 & 1 & \cite{kumi_high-speed_2010} \\
                        stage\\
			& & & \\
			Piezo stage & 0.03--0.09 & 0.28--1.5 & \cite{straub_near-infrared_2002} \\
			 & 0.06	& 0.065 & \cite{haske_65_2007} \\
			 & 10--30 & 1.5 & \cite{ovsianikov_laser_2011} \\
			& & & \\
			Galvo-galvo & 0.005--0.2 & 0.085--1.5 & \cite{thiel_direct_2010} \\
                        mirror & 0.01	 & 1.3 & \cite{maruo_three-dimensional_2000} \\
			 & 7 & 0.78--1 & \cite{farsari_two-photon_2006} \\
			 & 0.4--200 & 0.2--1.2 & \cite{obata_high-aspect_2013} \\
			& 21--103 & 0.086--0.43& \cite{gottmann_high_2009} \\
                        & 400 & 1--10 & \cite{SkylarScott2016MicrofabricatedBioscaffolds} \\
			& & & \\
			Rotating & 7200 & 1 & \cite{rensch_laser_1989} \\
                        polygon-galvo\\
                        mirror \\
			& & & \\
			Resonant-galvo & 3300--8200 & 1--4 & Present work \\
                        mirror
		\end{tabular}
	\end{center}
	\caption{Representative DLW laser scanning speeds and nominal
          minimum feature sizes reported in recent
          literature. ``Present work'' gives scan speed with the printer
          configured as described for most of the examples in this paper (1.6$\times$ zoom yielding a $416\times 416$-$\mu\textrm{m}$
          workspace, and printing during only the left-to-right sweep of the resonant scanner) and the maximum speed that we've used (bidirectional printing at 1.3$\times$ zoom). Minimum feature size and scan speed covary as described in the text.}
	\label{table:speed}
\end{table}

In a resonant-scanner--based system with a resonant frequency of $F_r$ and
useable workspace dimensions of $2\xi D$ along the scanning dimension,
the average beam speed is $2\xi DF_r$. For example, at
1.6$\times$ zoom, our system's useable workspace along the X axis
is approximately 412 $\mu$m, resulting in an effective mean beam
speed of 3.3 m/s. Note that this estimate assumes printing only in one direction of the laser scan
(\fig{fig:schematic}b); bi-directional printing effectively doubles
beam speed, although any misalignment of the two scan directions leads to inferior results.
Note also that decreasing magnification will increase the
distance that the beam travels while commensurately increasing
beam speed, leaving print time unchanged (provided that the laser can 
supply sufficient power to polymerise resist at the higher speed).

For our rDLW system, we can estimate fabrication time for an object from the linear 
printing distance (i.e., length of a scan line $2\xi D$ 
times the number of scan lines per layer $S_y$ times the number of layers $S_z$) times the 
mean beam speed. For the large block in \fig{fig:rulers}b, this results in an estimated 
fabrication time of $\sim 19$~s, which comports well with our actual print
time of $\sim 25$~s. DLW systems vary widely and there are no established benchmarks, 
making general comparisons of writing speed and printing time difficult \cite{LaFratta2007, Sun2004}. 
Galvanometer-based two-photon microscopes are typically an order of magnitude slower than resonant-scan 
microscopes. For example, at $512\times 512$ pixels, resonant-scan microscopes typically 
achieve 30-Hz frame rates while typical galvo-based systems achieve $\sim$1--2-Hz frame rates at
the same scan angle and resolution \cite{jonkman2015confocal}. A pure
galvanometer beam control system designed for calcium imaging might see a
beam speed of 200 mm/s \cite{obata_high-aspect_2013}. If such a system 
were used to write the simple block in \fig{fig:rulers}b, fabrication would take about 3.8 minutes.

We emphasise that these calculations are for an object with a fill ratio of 1 (i.e., 
100\% of the total object volume is polymerised), so these estimates represents 
a worst-case fabrication time for an object of this size. Objects with smaller
fill fractions---as would be likely for most objects of interest---would
see reduced fabrication times on galvanometer- or stage-based systems that optimise 
beam path to reduce total travel distance. As with estimates of accuracy and 
resolution, our estimates of printing speed are highly dependent on our choice of 
optical components, printing parameters, and 
photoresist. Significant improvements or diminishments in all assayed
metrics can be realised with a different choice of hardware,
laser power, or row/layer density.

\section{Discussion and Conclusion}
We reported on rDLW: a 3D printer based on a standard two-photon
microscope with a resonant raster scanner and our custom PrintImage control
application. The rDLW system provides several key features including full access to fabrication
parameters, high printing speeds,
and ease of extensibility. Building on the widely-used open-source ScanImage microscopy package, this work
provides a platform for future modifications and customisations.

Because our rDLW printer exposes all process parameters, and indeed all control
software, to the user, our system is easily
adaptable to experiments with novel fabrication techniques that take advantage of the unique
feature of voxel-by-voxel modulation of laser
power. This fine-grained power control proved useful in compensating for nonuniform optical effects such as vignetting, and could further be used to take advantage of intermediate states of polymerisation and the material
properties that so arise (i.e., refractive index, rigidity, or fluorescence)
\cite{Gissibl2016multilens}. In addition, laser power is nonlinearly correlated
with minimum feature resolution \cite{Kawata2001submicron}, so per-voxel power modulation could provide additional control of the sizes of different single-voxel features in a single print process.

The use of a tunable femtosecond laser adds significant cost to our system, and could be replaced
by fixed-wavelength fibre-based femtosecond sources. However, since tunable femtosecond lasers are common components
of two-photon microscopes, we suggest that this more flexible laser may open
up new material choices for polymerisation at a range of wavelengths.  We have demonstrated
the capabilities of the system using IP-Dip, a proprietary refraction-index--matched photoresist
designed for high-resolution two-photon polymerisation. However,
the wide tunable range of modern two-photon laser sources (for our laser, 700--1050 nm), or the ease with which another laser can be added to the beam path,
makes possible printing with commercial or custom resists having significantly different absorption
spectrum peaks. This capability would simplify fabricating compound structures
composed of multiple photoresists, each with different mechanical or optical
properties \cite{Zeng2015}.

A limitation of existing DLW techniques---which not infrequently
influences printed object design---is the need to add structural supports under
suspended features, lest gravity and movement of the photoresist during printing
displace the incomplete features before they are anchored to the body of the
object being printed. 
An unexpected benefit of the rDLW system is that the high speed
of printing allows, to some degree, the printing of unsupported, suspended features in viscous liquid photoresists (like
IP-dip). In addition to streamlining object design, the ability to print
without the need of support structure potentially enables the fabrication of
previously unrealisable objects.

Though the maximum print size of the described rDLW system ($\sim 400 \times 400 \times 350\;\mu\textrm{m}$) is suitable for many micro-scale applications,
there are use cases (e.g., tissue culture scaffolding) that require larger
object sizes while maintaining micron resolution. Several photoresists, including
IP-Dip, Ormocer, and SU-8, allow newly polymerised material to bond
directly onto previously polymerised material without mechanical defect. This allows
an object to be built by stitching together several overlapping sections,
each of which we refer to as a {\em metavoxel}. Additionally, whereas for a single metavoxel the zoom setting controls both X-axis resolution and maximum object size, stitching allows decoupling of these two parameters by printing a single
piece as multiple smaller overlapping pieces at higher magnification. When
stitching multiple metavoxels together, a stage with absolute
linear accuracy on the order of the desired resolution is required. While a discussion of stitching is outside the scope of this paper, we note that, as of this writing, PrintImage allows stitching using either the microscopy stage or a commercial hexapod system, thus allowing fast printing of arbitrarily large objects, and it can easily be extended to use other hardware.

While resonant scanners have been previously used in 2D laser printing
\cite{Schermer1990resonant, fujii1995scanning}, a more common approach
for raster-scan printing uses a multi-sided mirror rotating at constant speed to
sweep the across the workspace \cite{takizawa1997laser}.
Replacing the resonant scanner in our rDLW
printer with such a raster-scan mirror would triple print speed by eliminating the flyback and near-zero-speed portions of the beam path. It would allow nearly linear beam
speed, providing uniform voxel size and obviating sinusoidal
power compensation. Though this would remove the resonant
rDLW's capacity to increase print resolution without a concomitant reduction in
printing speed (i.e., zoom printing), a similar effect may be achievable by
incorporating a zoom lens. Conversely, a change in mirror rotation speed would
allow changes in X-axis resolution without affecting workspace size.

\section{Materials and Methods}
\subsection{Programming and analysis}
All programming, modeling and analysis was done in MATLAB (The
Mathworks, Framingham, MA) running under Windows 10 on a desktop
computer with an Intel i7 processor and 16 GB of RAM.
The PrintImage software is available at {\tt
  https://github.com/gardner-lab/printimage}. Its documentation lists its software
dependencies.

\subsection{Design of calibration objects and print models}
All custom benchmarking and example objects described here were
created with Solidworks2016 (Dassault Systèmes, Concord, MA) and
exported using the native STL converter. Calibration objects not
printed on our rDLW system (see Figure 2a) were printed using a Nanoscribe Photonic Professional GT (Nanoscribe GmbH, Stutensee, Germany). STL
files for the Darwin Bust (\fig{fig:schematic}d) and Torus Knot
(\fig{fig:logpile_and_knot}c) were obtained from the Museum of Applied
Arts and Sciences in Sydney, Australia and Tadej Skofic, respectively.

\subsection{Photoresists and Post-Processing}
A key step in developing a DLW solution is identifying photoresists that
are compatible with both the specifications of the printing process
(e.g., two-photon polymerisation, laser wavelength and power output, printing speed, etc.)
and the requirements of the application (e.g., hardness, adhesion,
biocompatibility, optical clarity, etc.). Though multiple photoresist formulations have
been described, the majority used for two-photon DLW consist of soluble organic
monomers or oligomers (typically acrylate derivatives) that are cross-linked,
and thus made insoluble, by free radicals or cations produce by the exposure
of a photoinitiator or photoacid generator \cite{Fourkas2015}. The use of
a tunable laser in the described rDLW system offers the possibility of printing with a
variety of commercially available (e.g., Ormocer, KMPR,
SU-8, etc.), custom, or proprietary photoresists.

In an effort to ensure that our assays were representative of the limits of
our rDLW system's performance, all objects reported here for illustration or benchmark measurements
were printed with a high-performance photoresist, IP-Dip (Nanoscribe GmbH, 
Stutensee, Germany). IP-Dip is a proprietary
liquid photoresist that is refraction-index--matched to glass to minimise optical
distortion and enable rapid, fine-resolution two-photon polymerisation. IP-Dip polymerises under 390-nm light (i.e., the two-photon 
effective wavelength of our 780-nm source), producing solid, semi-transparent acrylic
objects that have been used in biomedical, optical, and microfluidic applications.
Following printing, residual un-polymerised resist was removed by submerging the
substrate and printed element in a solvent, propylene glycol methyl
ether acetate (PGMEA), for approximately 20 min. The prints were then
rinsed in methoxy-nonafluorobutane\manufacturer{3M Novec 7100 
	Engineering Fluid} to remove trace PGMEA residue.

\subsection{Scanning Electron Microscopy}
Measurements of printed objects were made using SEM
micrographs. To enhance sample conductivity, the samples were
sputter-coated with gold\manufacturer{Sputter Coater 108} prior to
imaging. The samples were placed 3 cm under the gold target and were
coated for 1 min at 0.05 mbar and 20 mA. \manufacturer{SEM imaging
  was performed with the Supra 55VP Field Emission SEM (Zeiss).} The
samples were imaged at 6-mm working distance with the secondary
electron sensor, 3-kV accelerating voltage and 30-$\mu$m aperture
size.

\subsection{Energy Deposition Analysis}
When IP-Dip polymerises, its fluorescent intensity changes, allowing
printed objects to be imaged by exposing them to the laser at a power
that causes fluorescence but is below the polymerisation threshold. The
imaged fluorescent intensity is inversely proportional to the degree
of polymerisation.

The energy deposition profile was quantified by measuring the
fluorescent intensity of printed cubes at a depth of 5 $\mu$m below
their top surfaces. Images were made using the ScanImage software at 1$\times$ zoom, and
were analysed in MATLAB. As vignetting at the extreme corners of the
workspace reduces the laser's intensity beyond our ability to
compensate during printing, we restricted our analysis to objects up
to $400\times 400\;\mu\textrm{m}$ ($1.6\times$ zoom). To eliminate the effects of spatial
nonuniformity (such as vignetting) in our imaging system, rather than photographing still images and
measuring brightness values over the imaging plane, we instead moved
the printed objects under the lens (at 200 $\mu$m/s along the X axis) using our stitching
stage and recorded over time (at 15.21 Hz) the brightness values over the 1-pixel-by-10-$\mu$m XY region of the printed cube at the centre of the camera's reference frame
at each X-axis step. This was repeated for 15 equally spaced lines covering the Y axis.
In order to compensate for the
non--vignetting-corrected imaging laser power, a control image of the
field of view without any objects was captured, and intensity values of
the images of each test object were divided by the control image's
intensity.

\section{Author Contributions}
TJG and JMT conceptualised the project and implemented the initial
proof of concept. BWP wrote the software for controlling the print
process and integrated it with existing ScanImage routines. TMO and
JMT designed the microscope modifications to enable the new printing
process. CM developed test print objects, created calibration
protocols, made all SEM micrographs, and performed quantitative
testing. BWP, TMO, and CM jointly contributed to device and process
refinement. TJG and TMO supervised the project. BWP, CM, amd TMO
drafted the manuscript with input from all other authors.

\section{Acknowledgments}
We wish to thank 
Alice White (Boston University) for use of her
Nanoscribe Photonic Professional GT 3D printing system and her
expertise with microfabrication processes; Jacob Franklin (Vidrio
Technologies) for his assistance with ScanImage; Tadej Skofic for
designing the Torus Knot shown in \fig{fig:logpile_and_knot}; and
the Museum of Applied Arts and Sciences in Sydney, Australia for
providing the bust of Darwin shown in \fig{fig:schematic}. We also
wish to thank Alberto Cruz-Martin, Todd Blute, Ian Davison, Jeff Gavornik, 
and L.~Nathan Perkins for their helpful comments on
earlier drafts of this manuscript. Furthermore, we are grateful for the packages on the MATLAB File Exchange from Alex A (Mesh voxelisation), Eric Johnson (STL File Reader), Pau Mic\'o (stlTools), and Teck Por Lim (Significant Figures). This research was supported by NIH
grants (U01NS090454 and R01NS089679) and a sponsored research
agreement with GlaxoSmithKline.

\bibliography{3d-printing}

\end{document}